\documentstyle[11pt]{article}
\newcommand{\yourabstract}[1]{
\mbox{}\\
\mbox{}\\
{\bf\noindent Abstract}\\
\begin{center}
\mbox{}\parbox[t]{5.in}{#1}
\end{center} }
\begin{document}

\begin{titlepage}
\begin{flushright}
February 2001
\end{flushright}
\vskip 1.20cm
\begin{center}

{\bf \large Entropy Production in Relativistic Hydrodynamics}
\vskip 0.90cm

H.-Th. Elze$^{1,2}$, J. Rafelski$^1$ and L. Turko$^{1,3}$

\vskip 0.40cm $^1$ Department of Physics, University of Arizona,
Tucson, AZ 85721

\vskip 0.20cm $^2$ Instituto de Fisica, Universidade Federal do
Rio de Janeiro \\ C.P. 68528, 21945-970 Rio de Janeiro, RJ, Brasil

\vskip 0.10cm $^3$ Institute of Theoretical Physics, University of
Wroclaw \\ pl. Maksa Borna 9, 50-204 Wroclaw, Poland \\

\end{center}

\yourabstract{The entropy production occurring in relativistic
hydrodynamical systems such as the quark-gluon plasma (QGP) formed
in high-energy nuclear collisions is explored. We study mechanisms
which change the composition of the fluid, {\it i.e.}  particle
production and/or chemical reactions, along with chemo- and
thermo-diffusion. These effects complement the conventional
dissipative effects of shear viscosity, bulk viscosity, and heat
conductivity.}

\end{titlepage}

\noindent {\bf Introduction.} There is fundamental interest in the
study of high energy density matter, which promises to illucidate
further the properties of the strongly interacting vacuum and the
high temperature phases of strongly interacting matter \cite{Kar00,DeT95}.
The method of choice for the description of the rapidly
evolving and possibly inhomogeneous  quark-gluon plasma (QGP) has been
often the hydrodynamical approach \cite{Bjo83,Csernai}. It is generally
believed that QGP existed shortly after the Big Bang initial phase of the
universe, and that it may be re-created in relativistic nuclear
collisions \cite{QGP,CERN}.

In order to gain a better understanding of this new form of
matter, we explore here the entropy production due to numerous
dissipative effects arising in the evolution of QGP towards the
point of hadronization. The entropy content of the QGP phase at
hadronization determines the observable particle multiplicity
\cite{Let93}. Therefore, in order to establish via particle
multiplicity the initial conditions reached in heavy-ion
collisions \cite{Bjo83}, the effect of entropy production in QGP
evolution has to be understood.

To begin with, we present here the main result of this paper,
before embarking on its derivation. The entropy production is
given by:
\begin{eqnarray}
T\partial_\mu s^\mu &=&-\sum_i\mu_i{\cal J}_i
-T\sum_{i,j}\sigma_{ij}(\partial_\mu\frac{\mu_i}{T})\triangle^{\mu\nu}
(\partial_\nu\frac{\mu_j}{T}) \nonumber \\
&\;&-\frac{\kappa}{T}{\cal Q}_\mu\triangle^{\mu\nu}{\cal Q}_\nu
+\frac{\eta}{2}{\cal W}_\alpha^\beta\triangle_\beta^\gamma {\cal
W}_\gamma^\delta\triangle_\delta^\alpha +\zeta (\partial_\mu u^\mu
)^2 \;\;, \label{result} \end{eqnarray} see Eq.\,(\ref{28}). We
will show that the four-divergence of the entropy current,
$\partial_\mu s^\mu$, is always nonnegative and vanishes only,
when all space-time gradients of temperature $T$, of chemical
potentials $\mu_i$, and of four-velocity $u_\mu$ vanish, and
particle densities of species ``$i$'' assume their chemical
equilibrium values. -- The projector $\triangle$ is:
$\triangle_{\mu\nu}\equiv g_{\mu\nu}-u_\mu u_\nu$, with
$g_{\mu\nu}\equiv\mbox{diag}(1,-1,-1,-1)$ and $u^2=1$; we use
units such that $\hbar=c=k_B=1$.

The first two terms on the right-hand side of Eq.\,(\ref{result})
are new. They are due to the sources ${\cal J}_i$ of particle
production and chemo- and thermo-diffusion contributions involving
gradients of chemical potentials and temperature , respectively.
These terms arise in the continuity Eq.\,(\ref{6}) below; $\sigma$
denotes the matrix of mutual diffusion coefficients between the
chemical QGP components. Furthermore, the constants $\kappa$,
$\eta$, and $\zeta$ denote the transport coefficients of heat
conductivity, shear viscosity, and bulk viscosity, respectively.
The corresponding terms in Eq.\,(\ref{result}) involve the
heat-flow four-vector ${\cal Q}_\mu$, Eq.\,(\ref{24}), and the
shear tensor ${\cal W}_{\mu\nu}$, Eq.\,(\ref{25}), and have been
obtained earlier \cite{Csernai,Weinberg}.

Generally, dissipative processes are important, when space-time
gradients of hydro- or thermodynamical quantities in the system
become large relative to its relaxation scales. The latter and
especially the transport coefficients and source terms introduced
in Eq.\,(\ref{result}) have to be fit to experiment or calculated
from an underlying microscopic transport theory
\cite{Weinberg,Baym}. For a recent extensive review of such
state-of-the-art microscopic calculations and new gauge theory
results we refer to Ref.\,\cite{Arnold} and its numerous
references.

In the QGP case, particle production converting kinetic energy of
a nuclear collision into high multiplicity of new particles is a
characteristic feature, together with annihilation processes,
changing the composition of the plasma
\cite{Raf82,Kajantie1,Kajantie2}. Motivated by this fact, our aim
presently has been to incorporate these dissipative and entropy
producing mechanisms consistently into the relativistic
hydrodynamical framework.

\vskip .2cm \noindent {\bf Equations of Motion.} Presently we
consider the relativistic energy-momentum tensor for an imperfect
fluid:
\begin{equation}
T^{\mu \nu}\equiv (\epsilon +P)u^\mu u^\nu -Pg^{\mu \nu} +\delta
T^{\mu \nu}=T^{\nu \mu} \;\;, \label{1} \end{equation} where
$\epsilon$ and $P$ denote its local energy density and pressure,
respectively. The additional term $\delta T^{\mu \nu}$
incorporates the dissipative effects within the fluid
\cite{Csernai,Weinberg}. It will be obtained in detail in the
following. We point out that the corresponding transport
coefficients have to be determined consistently, {\it i.e.},
including also all contributions from particle production or other
composition changing processes, if present in the system, besides
the scattering contributions \cite{Baym,Arnold}.

The evolution of the hydrodynamical system then is governed by the
equation of motion,
\begin{equation}
\partial _\mu T^{\mu \nu} =f^\nu
\;\;, \label{2} \end{equation} where $f^\mu$ describes the density
of external forces, gravity for example, which act on the fluid
locally. A closed system has $f^\mu\equiv 0$, which results in
total four-momentum conservation. In order to obtain a complete
description of the fluid, the equation of motion (2) has to be
supplemented by an equation of state relating the energy density
to the pressure. In the simplest case of a noninteracting
ultrarelativistic system, we have $P=\epsilon /3$, while a
realistic QGP equation of state is given in
Ref.\,\cite{Ham00}.

Often, particularly if there exist conserved charges or particle
numbers, a simple functional relationship between pressure and
energy density does not exist. Then, assuming local thermal and
possibly chemical equilibrium, both can still be represented
parametrically in terms of a temperature $T=T(x)$ and possibly
several chemical potentials $\mu _i=\mu _i(x)$. In this case, one
has to complement Eqs.\,(\ref{1})-(\ref{2}) by continuity
equations,
\begin{equation}
\partial _\mu\rho _i^\mu\equiv \partial _\mu  \rho _iu^\mu =0
\;\;, \label{4} \end{equation} where $\rho _i\equiv\rho _i(x)$
denotes the particle number density for species ``$i$'' in the
local rest frame of a fluid cell. We will not distinguish the
cases of charge or particle number conservation and count each
(anti-)particle species separately. -- The local rest frame is
always determined by $u^\mu =u^\mu _{(0)}\equiv (1,0,0,0)$. Thus,
by definition, we have $\rho_{i(0)}^0=\rho _i$.

Similarly, the total energy density is defined to be equal to
$\epsilon$. Therefore, we must require that dissipative terms do
not contribute to $T^{00}$ in the local rest frame,
\begin{equation}
T_{(0)}^{00}=\epsilon \;\;\longrightarrow\;\;\; \delta
T_{(0)}^{00}=0 \;\;\longrightarrow\;\;\; u_\mu u_\nu\delta
T^{\mu\nu}=0 \;\;, \label{5} \end{equation} cf. Eq.\,(\ref{1}).
This constraint plays an important role in determining the general
structure of $\delta T^{\mu\nu}$ consistently with the Second Law
of Thermodynamics.

In order to incorporate diffusion and particle production and/or
chemical reactions among the constituents, which change the
composition of the fluid, we now generalize Eq.\,(\ref{4}) to
read:
\begin{equation}
\partial _\mu \tilde{\rho}_i^\mu \equiv
\partial _\mu (\rho _i^\mu +\triangle^{\mu\nu}
\partial _\nu {\cal R}_i)={\cal J}_i
\;\;, \label{6} \end{equation} with the new four-current
$\tilde{\rho}_i$ having the property $u_\mu\tilde{\rho}_i^\mu
=\rho _i$, since $u_\mu\triangle^{\mu\nu}=0$. Furthermore, ${\cal
J}_i$ denotes the local source for particle species ``$i$'', which
is a functional of the densities, ${\cal J}_i(x)\equiv{\cal
J}_i[\rho _j](x)$. It can be calculated in terms of microscopic
reaction cross sections. We do not consider here the possibility
of external sources. -- In a homogeneous system such continuity
equations become familiar chemical rate equations, which have
previously been applied to strongly interacting relativistic
systems on the hadronic and partonic level
\cite{Raf82,JanRep,Mueller}.

The explicit form of the diffusion term $\propto\partial {\cal R}
_i$ in Eq.\,(\ref{6}) will be determined consistently with the
Second Law of Thermodynamics. -- One might try to follow the
nonrelativistic ansatz which recovers the usual diffusive
contribution to the particle current in Eq.\,(\ref{6}) in the
local rest frame:
\begin{equation} \label{nrDiff}
{\cal R}_i\equiv\sum_j\tilde{\sigma}_{ij}\rho _j\;\;,\;\;\;
\vec{j}_i\equiv -\nabla {\cal R}_i
=-\sum_j\tilde{\sigma}_{ij}\nabla\rho _j \;\;,
\end{equation}
where $\tilde{\sigma}_{ij}$ represents the mutual diffusion
coefficient between species ``$i$'' and ``$j$''. However, this
ansatz turns out to be incompatible with the requirement that the
entropy never decreases and will be suitably generalized below.

\vskip .2cm \noindent {\bf Entropy Production in an Imperfect
Fluid.} Here we formulate a consistent theoretical framework to
study the entropy production due to dissipative effects within a
relativistic fluid. We assume local thermal equilibrium and
characterize the particle densities by appropriate chemical
potentials.

We remark that introducing chemical potentials does {\it not}
necessarily imply chemical equilibrium in the sense of saturating
the available phase space \cite{Jan}.
 -- For example, consider a thermalized electron-positron
plasma with one chemical potential regulating the charge density,
as usual, and  with a number of $e^+e^-$-pairs which does not
correspond to the given temperature. Such a situation can be
described by introducing a phase space occupancy factor in
addition to the charge chemical potential or, equivalently, by two
chemical potentials regulating the number of electrons and
positrons separately. -- We presently admit such cases, recall
Eq.\,(\ref{6}), which typically involve charge conservation side
by side with dynamically changing particle pair populations in the
fluid.

In order to obtain a relationship between the thermodynamical
quantities and variables discussed so far and the entropy, in
particular, we employ the (local) equilibrium relations,
\begin{equation}
-PV=\Omega (T,V,\mu _i)=U-TS-\sum_i\mu _iN_i
\;\;, \label{14}
\end{equation}
where  $U$, $S$, and $N_i$ denote the internal energy, entropy,
and particle numbers, respectively.

Two comments are in order here. Firstly, having nuclear collisions
in mind, we employ the grandcanonical description despite the fact
that the fireball created in a single event is thermodynamically
closed. There is no heat or particle bath. However, instead of
generating the grandcanonical ensemble by our system being
connected to such a bath, we consider a large number of events to
form the ensemble. The events have to be identical with respect to
the conserved total energy-momentum, angular momentum, and baryon
number, to name the most relevant ``macroscopic'' observables.
Nevertheless, microscopically, the initial state may differ from
event to event. Then, under the assumption that the entropy is
maximized, the grandcanonical ensemble is obtained
\cite{Jan,LandauLif}. In the hydrodynamical approach we apply this
reasoning locally to each fluid cell.

Secondly, consider finite size effects present in small volume
systems. In this case, the usual ($V\rightarrow\infty$)
statistical description breaks down, if the unphysical
fluctuations of exactly constrained quantities are not
eliminated.\footnote{Such constraints arise due to exact
conservation laws, unbroken internal symmetries, and quantization
of the motion of the constituents (see, e.g.,
Refs.\,\cite{DanosJan,Turko,Elze,Ben}). For example, the thermodynamical
properties of small quark-gluon plasma  droplets start to deviate
strongly from the asymptotic ($V\rightarrow \infty$) behavior, if
the dimensionless parameter $TV^{1/3}$ becomes of order one or
less.} This issue needs to be addressed in our context in the
future.

We also need the Duhem-Gibbs relation for the densities
corresponding to Eq.\,(\ref{14}),
\begin{equation}
Ts=\epsilon +P-\sum_i\mu _i\rho _i \;\;. \label{15}
\end{equation}
Then, using Eqs.\,(\ref{1}) and (\ref{15}), we obtain by
contraction with $u_\nu$ of the equation of motion (\ref{2}):
\begin{eqnarray}
u_\nu f^\nu&=&u_\nu\partial _\mu T^{\mu\nu}=u_\nu\partial _\mu
([Ts+\sum_i\mu _i\rho _i]u^\mu u^\nu)-u^\mu\partial _\mu P
+u_\nu\partial _\mu\delta T^{\mu\nu}
\nonumber \\
&\;&=u_\nu u^\nu\partial _\mu [T\tilde{s}^\mu +\sum_i\mu _i\rho
_i^\mu ] +[T\tilde{s}^\mu +\sum_i\mu _i\rho _i^\mu ]u_\nu\partial
_\mu u^\nu -u^\mu\partial _\mu P +u_\nu\partial _\mu\delta
T^{\mu\nu}
\nonumber \\
&\;&=\partial _\mu [T\tilde{s}^\mu +\sum_i\mu _i\rho _i^\mu
]-u^\mu\partial _\mu P +u_\nu\partial _\mu\delta T^{\mu\nu} \;\;,
\label{16} \end{eqnarray} where we introduced an entropy
four-current $\tilde{s}^\mu\equiv su^\mu$ and the currents $\rho
_i^\mu\equiv\rho_iu^\mu$, as before. For the last equality, we
made use of $u^2=1$.

For conservative external forces which can be derived from a
potential (density), $f^\mu\equiv\partial^\mu\phi$, using
Eq.\,(\ref{6}) and $u^\mu\partial _\mu  =d/d\tau$, we can further
simplify Eq.\,(\ref{16}) ($\dot{A}\equiv dA/d\tau$):
\begin{eqnarray}
0&=&T\partial _\mu\tilde{s}^\mu + \sum_i\mu _i({\cal J}_i
-\partial _\mu \triangle ^{\mu\nu}\partial _\nu {\cal R}_i)
+s\dot{T}+\sum_i\rho _i\dot{\mu} _i-\dot{P}-\dot\phi
+u_\nu\partial _\mu\delta T^{\mu\nu}
\nonumber \\
&=&T\partial _\mu\tilde{s}^\mu +\sum_i\mu _i({\cal J}_i-\partial
_\mu \triangle ^{\mu\nu}\partial _\nu {\cal R}_i) +u_\nu\partial
_\mu\delta T^{\mu\nu} \;\;, \label{17} \end{eqnarray} where the
terms involving $d/d\tau$ cancel identically. In order to
demonstrate this cancellation, we multiply them by $V\cdot d\tau$
and employ Eq.\,(\ref{14}),
\begin{eqnarray}
&\;&SdT+\sum_iN_id\mu _i-VdP-Vd\phi
\nonumber \\
&=&d(TS)+d(\sum_i\mu _iN_i)-d(PV)-TdS-\sum_i\mu _idN_i+PdV-Vd\phi
\nonumber \\
&=&dU-TdS-\sum_i\mu _idN_i+PdV-Vd\phi =0 \;\;. \label{19}
\end{eqnarray} The last equality expresses the First Law of
Thermodynamics, which is incorporated here.

Finally, we introduce the proper entropy four-current defined by:
\begin{equation}
s^\mu\equiv\tilde{s}^\mu +T^{-1}u_\nu\delta T^{\mu\nu} +\sum_i{\cal
L}_i\triangle ^{\mu\nu}\partial _\nu {\cal R}_i \;\;. \label{20}
\end{equation} This form is found by `trial and error'
and, by the constraint (\ref{5}) and $u_\mu\triangle^{\mu\nu}=0$,
implies $s_{(0)}^0=s$ in the local rest frame.

We remark that the difficulty to identify a viable form of the
entropy four-current, {\it i.e.} the admissible structure of
$\delta T^{\mu\nu}$, ${\cal L}_i$, and ${\cal R}_i$, resides in
the requirements that the entropy  density be equal to $s$ in the
local rest frame and that it never decreases anywhere in the
system. In the following we will determine the function ${\cal
L}_i$ consistently together with ${\cal R}_i$ and $\delta
T^{\mu\nu}$, in order to obtain the new contributions due to
diffusion and, particularly, due to particle production, which
have not been considered before.

We combine Eqs.\,(\ref{17}) and (\ref{20}), which yields:
\begin{eqnarray}
T\partial _\mu s^\mu &=&-\sum_i\mu _i{\cal J}_i-T^{-1}(\partial
_\mu T)u_\nu \delta T^{\mu\nu} +(\partial _\mu u_\nu )\delta
T^{\mu\nu}
\nonumber \\
&\;&+\sum_i(\mu _i+T{\cal L}_i)\partial _\mu\triangle
^{\mu\nu}\partial _\nu  {\cal R}_i +T\sum_i(\partial _\mu{\cal
L}_i)\triangle ^{\mu\nu}\partial _\nu  {\cal R}_i \;\;. \label{21}
\end{eqnarray} This equation describes the entropy production in
an imperfect fluid evolving under the influence of conservative
external forces, which do not generate entropy (Liouville's
theorem). In order to agree with the Second Law of Thermodynamics
for the closed system ($f^\mu\equiv 0$), the right-hand side of
Eq.\,(\ref{21}) has to be nonnegative for all fluid
configurations, such that $\partial _\mu s^\mu\geq 0$ always. This
requirement is implemented in the following steps.

First, we put a constraint on the sources introduced in
Eq.\,(\ref{6}),
\begin{equation}
\sum_i\mu _i{\cal J}_i\leq 0 \;\;. \label{22} \end{equation} This
constraint and its implications will be further discussed below,
beginning with Eq.\,(\ref{a2}). Under fairly general assumptions
about the particle producing processes or chemical reactions the
inequality (\ref{22}) guarantees that the system is driven towards
chemical equilibrium.

Second, the structure of $\delta T^{\mu\nu}$ including terms of
first order space-time gradients, which is subject to the
constraint (\ref{5}) and compatible with $\partial _\mu s^\mu\geq
0$, is well known \cite{Weinberg}:
\begin{equation}
\delta T^{\mu\nu}=\kappa (\triangle ^{\mu\gamma}u^\nu +\triangle
^{\nu\gamma} u^\mu ){\cal Q}_\gamma +\eta\triangle
^{\mu\gamma}\triangle ^{\nu\delta} {\cal W}_{\gamma \delta}
+\zeta\triangle ^{\mu\nu}\partial _\gamma u^\gamma \;\;,
\label{23} \end{equation} where $\kappa,\,\eta,\,\zeta\geq 0$
denote the coefficients of heat conductivity, shear viscosity, and
bulk viscosity, respectively, and with the heat-flow four-vector
${\cal Q}_\mu$ and the shear tensor ${\cal W}_{\mu\nu}$ defined
by:
\begin{eqnarray}
{\cal Q}_\mu&\equiv&\partial _\mu T-Tu^\nu\partial _\nu u_\mu
\;\;, \label{24} \\
{\cal W}_{\mu\nu}&\equiv&\partial _\mu u_\nu +\partial _\nu u_\mu
-\frac{2}{3} g_{\mu\nu}\partial _\gamma u^\gamma
\;\;. \label{25} \end{eqnarray}

Third, as we shall see in Eq.\,(\ref{28}), the contribution to the
right-hand side of Eq.\,(\ref{21}) of the diffusion terms
involving ${\cal R}$ becomes a nonnegative quadratic form if we
identify:
\begin{equation}
{\cal L}_i\equiv -\frac{\mu _i}{T}\;\;,\;\;\;  {\cal
R}_i\equiv\sum_j\sigma _{ij}\frac{\mu _j}{T} \;\;, \label{26}
\end{equation} where the symmetric matrix $\sigma$ consists of the
mutual diffusion constants and supposedly has only nonnegative
eigenvalues. We remark that ${\cal R}_i$ as defined in (\ref{26})
yields a diffusive particle current in the local rest frame,
\begin{equation}
\vec{j}_i\equiv -\nabla {\cal R}_i=-\sum_j\frac{\sigma
_{ij}}{T}(\nabla\mu _j-\frac{\mu _j}{T}\nabla T) \;\;. \label{27}
\end{equation} Thus, we find here a combination of chemo- and
thermo-diffusion contributions, which appears to be unique up to
terms involving first-order space-time gradients.

We observe that the diffusion current of Eq.\,(\ref{27}) indeed
generalizes the nonrelativistic ansatz discussed before, cf.
Eqs.\,(\ref{nrDiff}). For example, neglecting interactions, a
straightforward calculation in the ultrarelativistic Boltzmann
approximation gives:
\begin{equation}
\mu _j=T\mbox{ln}[C\rho _jT^{-3}] \;\;, \label{B1} \end{equation}
where $C$ collects the constants. Using this in Eq.\,(\ref{27}),
{\it i.e.} $\mu _j=\mu _j(\rho _j,T)$, we obtain:
\begin{equation}
\vec{j}_i=-\sum_j\sigma _{ij}(\frac{\nabla \rho _j}{\rho _j}
+3\frac{\nabla T}{T}) \;\;. \label{B2} \end{equation} Obviously
the term $\propto\nabla\rho _j$ corresponds to the nonrelativistic
one, while our covariant considerations additionally yield an
unique thermo-diffusion term.

Collecting the results of this section, we can now rewrite
Eq.\,(\ref{21}) with all the sources of entropy production in an
explicitly nonnegative form:
\begin{eqnarray}
T\partial _\mu s^\mu &=&-\sum_i\mu _i{\cal J}_i- T\sum_{i,j}\sigma
_{ij}^{-1}(\partial _\mu {\cal R}_i) \triangle ^\mu _\lambda
(\partial _\nu {\cal R}_j) \triangle ^{\nu\lambda}
\nonumber \\
&\;&-\frac{\kappa}{T}{\cal Q}_\mu \triangle ^\mu _\lambda {\cal
Q}_\nu \triangle  ^{\nu\lambda}+\frac{\eta}{2} \triangle _\alpha
^\beta {\cal W}_\beta ^\gamma \triangle _\gamma ^\delta \triangle
_\delta ^\mu {\cal W}_\mu ^\nu \triangle _\nu ^\alpha  +\zeta
(\partial _\mu u^\mu)^2 \;\;, \label{28} \end{eqnarray} which is
obtained with the help of Eqs.\,(\ref{23})-(\ref{26}) and by
making use of the properties of the projector $\triangle$. All
transport coefficients are treated as constants here, despite the
fact that they implicitly depend on the thermodynamical variables,
if calculated microscopically \cite{Arnold}.

Together with condition (\ref{22}), the Eq.\,(\ref{28}) guarantees
that a closed system evolves from local towards global
equilibrium. The latter is reached, of course, when all space-time
gradients of $T$, $\mu _i$, and $u^\mu$ vanish and the particle
densities have their asymptotic values, such that the sources
${\cal J}_i$ vanish.

\vskip .2cm \noindent {\bf Driving Towards Chemical Equilibrium.}
We now turn to the discussion of the local source terms ${\cal
J}_i[\rho _j]$, which describe the particle production in the
continuity equations (\ref{6}). This will clarify the meaning of
the constraint (\ref{22}).

We assume a sufficiently small fluid cell with a given volume,
such that all spatial gradients can be neglected in
Eq.\,(\ref{6}). This does not limit our considerations which are
related to the invariant source terms ${\cal J}_i$. Thus, in the
local rest frame, we obtain approximately:
\begin{equation}
\partial _t\rho _i={\cal J}_i\equiv
\sum _j[G_{i\leftarrow j}\rho _j-L_{j\leftarrow i}\rho _i] \;\;,
\label{a2} \end{equation} where the source term is expressed in
terms of gain and loss rates, $G$ and $L$, respectively. In
general, they are complicated  function(al)s of the temperature
and the densities themselves (see, for example,
Refs.\,\cite{Raf82,JanRep,Mueller} and Chap. X. in
Ref.\,\cite{LandauLif}).

The master equations (\ref{a2}) form a first order nonlinear flow
system \cite{Schuster}. In the following we will show that
Eq.\,(\ref{22}) presents a sufficient condition for an attractive
fixed point of the flow, which corresponds to the local chemical
equilibrium.

In order to proceed, we introduce the equilibrium densities
$\bar{\rho}_i(T)\equiv\rho_i(\bar\mu_i,T)$, where $\bar\mu_i$
denotes the corresponding equilibrium chemical potential, and the
reduced densities $\xi _i\equiv\rho _i/\bar{\rho}_i$, $0\leq\xi
_i<\infty$. We assume that
\begin{equation} \mu _i(\xi_i,T)\equiv\bar\mu_i+f_i(\xi_i,T)
\;\;, \label{a3}
\end{equation}
with $f_i(1,T)=0$, is a strictly increasing function of $\xi _i$
at a given temperature $T$. This is exemplified by the result
obtained in the Boltzmann approximation, Eq.\,(\ref{B1}),
\begin{equation}
\mu _i(\xi_i,T)=\bar\mu_i+T\mbox{ln}[\xi _i] \;\;. \label{a4}
\end{equation} In any case, the point
$\vec{\xi}_1\equiv\{\xi _i=1,$ for all $i\}$ in the
multidimensional $\xi$-space corresponds to the chemical
equilibrium state.

Sufficiently close to equilibrium, $\vec{\xi}\approx\vec{\xi}_1$,
the constraint (\ref{22}) can be expanded. We suppress the
temperature dependence from now on, tacitly assuming (as before)
that local thermal equilibrium is maintained by sufficiently fast
elastic scattering among the constituents of the considered fluid
cell. For convenience, we introduce rates weighted by the
equilibrium densities, $\bar{G}_{i\leftarrow j}\equiv
G_{i\leftarrow j}\bar{\rho _j}$ and $\bar{L}_{j\leftarrow i}\equiv
L_{j\leftarrow i}\bar{\rho _i}$. Then, employing Eq.\,(\ref{a3})
and using principle of detailed balance,
\begin{equation}
\sum _j[\bar{G}_{i\leftarrow j}(\vec{\xi}_1) -\bar{L}_{j\leftarrow
i}(\vec{\xi}_1)]=0 \;\;, \label{a5} \end{equation} {\it i.e.}
${\cal J}_i=0$ in chemical equilibrium, we obtain:
\begin{equation}
0\leq -\sum _i\mu _i{\cal J}_i\approx -\sum _{i,j}\bar\mu_i
(\vec{\xi}-\vec{\xi}_1)\cdot\nabla(\bar{G}_{i\leftarrow j}\xi _j
-\bar{L}_{j\leftarrow i}\xi _i)  |_{\vec{\xi}_1} \;\;. \label{a6}
\end{equation} The first nonvanishing contribution here is a
positive gradient term; in the absence of equilibrium chemical
potentials one obtains a curvature term here instead. Therefore,
the nonnegative contribution $-\sum _i\mu _i{\cal J}_i$ to the
entropy production, Eq.\,(\ref{28}), vanishes in the chemical
equilibrium state.

In order to prove that the equilibrium point $\vec{\xi}
=\vec{\xi}_1$ is actually an attractive fixed point, we suitably
rewrite the flow system, Eqs.\,(\ref{a2}),
\begin{equation}
\partial _t(\xi _i-1)=\bar{\rho}^{\;-1}_i
\sum _j[\bar{G}_{i\leftarrow j}\xi _j-\bar{L}_{j\leftarrow i}\xi _i]
\;\;, \label{a7} \end{equation}
and introduce a distance measure ${\cal M}$,
\begin{equation}
{\cal M}(\vec{\xi},\vec{\xi}_1)\equiv\frac{1}{2}\sum _i
\bar{\rho}_if'_i(1)
(\vec{\xi}-\vec{\xi}_1)_i(\vec{\xi}-\vec{\xi}_1)_i \geq 0  \;\;,
\label{a8} \end{equation} since $f'_i\equiv df _i/d\xi _i>0$, $\mu
_i$ being a strictly increasing function. Then, we obtain:
\begin{equation}
\partial _t{\cal M}(\vec{\xi},\vec{\xi}_1)=\sum _i(\mu _i(\xi
_i)-\bar\mu_i)\bar{\rho}_i\partial _t(\xi _i-1)=\sum _i\mu _i(\xi
_i){\cal J}_i(\vec{\xi}) \leq 0 \;\;, \label{a9} \end{equation}
which holds sufficiently close to the equilibrium state, where
$\mu _i(\xi _i)-\bar\mu_i\approx f'_i(1)(\xi _1-1)$. We also made
use of Eqs.\,(\ref{22}), (\ref{a2}), and (\ref{a7}). Thus, the
system evolves towards the chemical equilibrium fixed point,
provided the constraint (\ref{22}) is satisfied.

In order to arrive at the second equality in Eq.\,(\ref{a9}), we
had to take into account charge and/or particle number
conservation. The equilibrium values of the chemical potentials
$\bar\mu_i$ are given by the equilibrium value of the chemical
potentials associated to conserved quantities. In the case of a
single charge conservation law this is written as:
\begin{equation}
\bar\mu_i=q_i\bar\mu\;\;,
\end{equation}
where $q_i$ denotes the generic charge of (anti-)particles of
species ``$i$''. Thus we obtain the relation:
\begin{equation}\label{charge}
\sum_i\bar\mu_i\bar\rho_i\partial_t\xi_i
=\bar\mu\partial_t\sum_iq_i\rho_i =0 \;\;, \end{equation} which we
employed in Eq.\,(\ref{a9}). The generalization including several
distinct abelian conservation laws is obvious, while the
consideration of nonabelian charges in this context presents an
open problem, cf. Ref.\,\cite{Tur00}.

In passing we mention that
\begin{equation}
\tau ^{-1}\equiv -\partial _t\mbox{ln}[{\cal
M}(\vec{\xi},\vec{\xi}_1)] \;\;, \label{a10} \end{equation} may
provide a useful measure of the chemical equilibration rate.
According to Eq.\,(\ref{a9}), it can be written as a sum of
partial equilibration rates, $\tau ^{-1}=\sum _i\tau _i^{-1}$,
with $\tau _i^{-1}\propto -\mu _i{\cal J}_i$, while a more
explicit form of the source terms, cf. Eq.\,(\ref{a2}), allows to
relate it to the equilibration times of chemical rate equations.

\vskip .2cm \noindent {\bf Discussion of the First-order
Formalism.} Some general remarks about the first-order dissipative
relativistic fluid dynamics presented in this work are in order
here. It is labeled ``first order'', since the entropy current,
Eq.\,(\ref{20}) together with Eqs.\,(\ref{23})-(\ref{26}),
contains only up to first order gradients, which describe
deviations from the equilibrium state of the fluid.

It has been argued before that in a wide class of first-order
theories the equilibrium state is unstable under physically
admissible perturbations \cite{Hiscock}. This includes the
theories of Eckart \cite{Eckart} and of Landau and Lifshitz
\cite{LandauLif2} especially. Unrealistically large growth rates
of the unstable modes\footnote{However, these were calculated only
for macroscopic fluids, such as water. It is not obvious that the
instabilities would be unrealistically fast for fluid parameters
representing microscopic fluids, such as QGP droplets.} led the
authors of Refs.\,\cite{Hiscock} to discard these theories in
favor of the Israel-Stewart second-order theory \cite{Israel},
where only stable or damped modes are found. The expense is the
increased complexity of the theory based on a larger number of
degrees of freedom. This goes together with an increased number of
phenomenological parameters, which ultimately should be calculated
microscopically, as we discussed.

We do not know, whether the first-order approach presented here
prevents such unphysical instabilities. A full linear response
analysis studying this problem is beyond the scope of this paper
and is left for future work.

However, the growth of potentially unstable modes will be cut off
by the finite size of the system in most realistic applications
and by the additional dissipative processes taking place at the
surface, such as radiation losses. These can be naturally
incorporated in the future. Furthermore, we do take into account
that the fluid may undergo dynamical changes of its composition.
This is particularly important for relativistic fluids at extreme
energy densities, where particle production is an important
dissipative process, as long as chemical equilibrium is not fully
established. Our considerations related to particle sources have
shown that small perturbations of the chemical equilibrium are
damped and decay. The system is driven by chemical reactions or
particle production towards the equilibrium state.

\vskip .2cm \noindent {\bf Conclusions.} We have presented here
the theoretical considerations which allow to determine the
entropy production which is consistent with the restriction to
first-order space-time gradients of the thermodynamical and fluid
variables, as well as with the requirements of the First and
Second Laws of Thermodynamics. Our result in Eq.\,(\ref{result})
and the subsequent hydrodynamical framework are generally
applicable to special relativistic fluids. As new elements, which
need to be considered in applications to high-energy nuclear
collisions and QGP, we introduce here the entropy producing
mechanisms of chemo- and thermo-diffusion together with the
contribution of particle production.

We point out that nonlinear flow systems, such as the one
describing the particle production processes in QGP, which we
studied in relation to chemical equilibration, generally tend to
show more structure than a simple attractive fixed point
\cite{Schuster}. Therefore, one may speculate, whether some
particular fluctuations may be caused by these nonlinearities in
high-energy collisions (high multiplicity events), {\it i.e.} by
the nonequilibrium chemistry of the reactions, with interesting
consequences for the entropy production.

Summarizing, the Eqs.\,(\ref{result}), (\ref{2}), and (\ref{6}),
together with the relevant definitions and supplemented by an
equation of state, present a consistent framework for the study of
dissipative relativistic hydrodynamical systems. This comprises
imperfect fluids with composition changing processes, particle
production in particular, and the more familiar dissipative
effects of heat conduction, shear and bulk viscosities, and
diffusion. We discussed the potential shortcomings of this
formalism, which involves only up to first-order space-time
gradients.

\vskip .2cm \noindent {\it Acknowledgements:} Work supported in
part by a grant from the U.S. Department of Energy,
DE-FG03-95ER40937\,, by PRONEX (Brazil) (No. 41.96.0886.00), and
by the Polish Committee for Scientific Research under contract
KBN-2\,P03B\,030\,18.


\end{document}